\documentstyle [12pt]{article}
\newcommand{\be}{\begin{equation}}
\newcommand{\ee}{\end{equation}}
\newcommand{\ra}{\rightarrow}
\newcommand{\gsim}{\stackrel{>}{\sim}}
\newcommand{\lsim}{\stackrel{<}{\sim}}

\begin{document}

\begin{titlepage}
\begin{flushright}
 IFUP-TH 6/94\\
 January 1994\\
  gr-qc/9401027
\end{flushright}

\vspace{7mm}

\begin{center}

{\Large\bf  Black Holes as Quantum Membranes}\\

\vspace{12mm}
{\large Michele Maggiore}

\vspace{3mm}

I.N.F.N. and Dipartimento di Fisica dell'Universit\`{a},\\
piazza Torricelli 2, I-56100 Pisa, Italy.\\
\end{center}

\vspace{4mm}

\begin{quote}
\hspace*{5mm} {\bf Abstract.} We propose a quantum description of black
holes. The  degrees of freedom to be quantized are
identified with the microscopic degrees of freedom of the horizon, 
and their dynamics is governed by the action of the relatistic bosonic
membrane in $D=4$. We find that a consistent and plausible description
emerges, both at the classical and at the quantum level. We present
results for the level structure of black holes. We find a
``principal series'' of levels, corresponding to quantization of the
area of the horizon. From each level of this principal series starts a
quasi-continuum of levels due to  excitations of the membrane.
We discuss the
statistical origin of the black hole entropy and the relation with
Hawking radiation and with the information loss problem.
The limits of validity of the membrane approach
turn out to coincide with the known  limits of validity of
the thermodynamical description of black holes. 

\end{quote}
\end{titlepage}
\clearpage

\section{Introduction}

In the search of a quantum theory of gravitation
black holes might play a role similar to the hydrogen atom 
in quantum mechanics: they might be the system in which important
conceptual problems and possibly even contradictions arise
in a sharp form, and from which we can start to understand what are 
the ``rules of the game'' when both quantum mechanics
and general relativity play an important role. 
An important problem
 arising in this context is the information loss paradox
raised by Hawking~\cite{Haw}, which apparently
 suggests that the evolution
of states in a black hole background violates such a basic principle
as unitarity. A closely related issue is the origin of the black
hole entropy: the relation between the entropy of a black hole  and
the area of the horizon, $S=A/4$ in Planck units,
 suggests the existence of a large number of
microstates (on the order of
$\sim\exp\{ 4\pi M^2\} $ for a Schwarzschild black
hole of mass $M$),
of which there is no trace in the classical description.
Another, less discussed, but possibly very important
issue is the relation between black holes and elementary
particles; 't~Hooft~\cite{tH}  has proposed that the two concepts 
should  merge as we cross the Planck mass (see also~\cite{HW}).

The hearth of the problem is that we do not know what is a black
hole at the quantum level. Indeed, it is not even clear what 
are the degrees of freedom to be quantized. Various authors,
from early work of Bekenstein~\cite{Bek} to more recent works
of 't~Hooft~\cite{tH} and Susskind, Thorlacius and Uglum~\cite{Sus1}
have suggested that the black hole is a normal quantum system with
discrete energy levels. But it is far from clear what is the
``Schroedinger equation'' for a black hole, if any,
or whether it can be
described by a wave function.

A first important hint for understanding  what are the 
degrees of freedom to be quantized comes from the membrane 
paridigm~\cite{TPM}, 
in which the interaction of a classical black hole with the
external environment, as seen by fiducial observers\footnote{We will 
hereafter refer to observers external to 
the black hole and static with
respect to it as ``fiducial'' observers. The precise definition of
fiducial observers for both non-rotating and rotating black holes is
given in ref.~\cite{TPM}.},
is entirely  described in terms of a fictitious membrane located 
close to the horizon and endowed with physical properties like
electric conductivity, temperature and viscosity.

At the quantum level, the description of the black hole 
horizon as a quantum surface with approximately one degree of freedom
per Planck unit of area has been 
advocated by 't Hooft~\cite{tH,tH2}, and it is strongly suggested by
the proportionality between the entropy and the area of the horizon.

A possible 
 objection to such a description is that, although from the
point of view of a fiducial observer the black hole behaves {\em as
if\/} there were a membrane on the horizon, such a membrane certainly
does not exist for the free falling observer. Then, it is not clear
what is the meaning of assigning actual microscopic degrees of freedom
to the membrane. Furthermore,
in order to have a useful description
it turned out~\cite{TPM} 
to be necessary to move the membrane a bit outside the
horizon, defining the so-called stretched horizon.
This allows to get rid of the details of the infalling matter
and  regularizes  the infinite red-shift factor
between the horizon and infinity. The amount of stretching is however
quite arbitrary, so that the position of the membrane
is determined
by our ``regularization'' procedure, rather than  by physics.  

The answer to the first objection follows from the recent 
works~\cite{Sus1,Sus2,Sus3,Sus4}, where a principle of ``black hole
complementarity'' is proposed. The principle states that the points
of view of the free falling and fiducial observers are complementary,
in a sense close to Bohr complementarity:
even if an ideal ``superobserver'',  who can compare the
measurements of both the fiducial and the free falling observers, 
would conclude that their observations
  are actually contradictory, still no
real  contradiction arises because such superobservers cannot be
realized in Nature.
More precisely, the analysis of several Gedanken
experiments~\cite{Sus3} indicates that 
any attempt to compare the results of the measurements
of a free falling observer crossing the horizon and of a fiducial
observer turns out to involve  assumptions on physics beyond
the Planck scale.
Similar considerations have been made in ref.~\cite{tH3}, where it is
argued that the commutators of the operators measuring Hawking
radiation with operators inside the horizon grow uncontrollably
large and prevent simultaneous measurements of the relative
observables. This suggestion has been elaborated in the very recent
paper~\cite{SVV}. 

This removes the first objection to the membrane approach.
The membrane actually {\em exists} for the fiducial
observer, who can legitimately assign  it microphysical degrees of
freedom 
and try to study the dynamics of this microscopic structure. The
membrane is not detected by the free falling observer who is crossing
the horizon, but this does not lead to any logical contradiction.

The second objection, concerning the arbitrariness in the position 
of the stretched horizon,  can  be  answered taking
into account the generalized uncertainty principle in quantum gravity,
which assigns to the horizon a physical thickness because of quantum
fluctuations~\cite{MM}. 

The principle of black hole complementarity has deep physical
consequences which imply a radical change in
our view of space-time.
In particular, as discussed by  Susskind~\cite{Sus4}, it
implies that even 
the notion of invariant event cannot be anymore relied upon. 
It seems to us that, if the ideas 
suggested by 't~Hooft and by Susskind and collaborators 
will prove to be correct, they
will represent a major 
conceptual advance in our understanding of quantum gravity. 

If we accept the point of view of black hole complementarity,
the membrane is the natural  candidate
for the location of the microstates of the black hole
and the
degrees of freedom of the membrane are interpreted
as the variables to be quantized.  

In this paper, in order to test these ideas, we carry them a bit
further and propose a specific dynamical framework for the membrane.
More precisely, we wish to explore the consequences
of the following assumption: from the point of view of a fiducial
observer, a black hole is described  by a membrane whose dynamics is
governed by an action principle. The action is taken to be that
of the closed relativistic bosonic membrane in $D=4$ dimensions,
which is proportional to the world-volume swept.

In sect.~2 
we will show that this description makes sense at the classical
level. While it is known that no stable classical solution of this
action exists in flat space, we will find that in the external,
fixed, background of
a black hole (or, more in general, in metrics with a horizon, as in
Rindler space)
there is a stable classical solution which lies just
on the horizon and other classical solutions, whose motion is confined
to the region outside the horizon and which approach the horizon in a
logarithmically divergent time.
This result, independently of the
applications to black hole physics, might be of interest 
in itself in the study
of relativistic membranes.

We will then proceed to a quantum description (sect.~3). 
At this stage, we will
have to resort to an  approximation in which the
only degree of freedom retained is the  radius of the membrane,
which is taken to have spherical symmetry.  Some 
aspects of black hole physics, like the value of
the entropy, are irretrievably lost in this approximation, since we
are drastically reducing the number of degrees of freedom;
however, many
interesting features survive, and therefore it is a useful
starting point, at least for qualitative understanding. The great
advantage of the approximation is that the problem is
reduced to an ordinary problem of quantum mechanics, rather than of
quantum field theory. 
We will find that an interesting and plausible quantum
structure emerges, both for Schwarzschild and 
more general Reissner-Nordstrom black
holes.  In the Rindler case, we will find
that the dynamics of the membrane is governed by
Liouville quantum mechanics.
We will discuss the limits of validity of the membrane approach
(sect.~\ref{limit}) and, remarkably, we will find that they coincide
with the limits of validity of the thermodynamical description of
black holes found in ref.~\cite{PSSTW}.

Using these results, in sect.~4 we will be able to propose an energy
level structure for black holes.

We will be mainly concerned with static properties, in particular with
the energy level structure; however, we will also make some comments
on the transition amplitudes between these levels, and their relation
with Hawking radiation (sect.~5).

It should be stressed at this point that, while the mathematical
formalism that we use
is well-defined, the physical interpretation of our results,
at the present stage, is only tentative. This is  unavoidable in
a situation in which we are trying to unravel what the ``rules of the
game'' are, and what physical concepts are appropriate for a quantum
description of black holes.  Thus, our results can only be considered
a possible starting point for further investigations.

At the appropriate points we will discuss the limits of validity of
our computations. However, we want to stress now that the membrane
approach is not, at our present level of understanding, a
fundamental description of black holes. We rather think to it as an
effective description, valid in a domain to be determined below.
In some sense, the situation is similar to the large distance effects
in {\em QCD}, where the quark-antiquark potential at scales of order
of 1~fm is described, at the effective level, in terms of a string.
Of course, this string is a useful description only on a given scale
of distances: if we look more closely, there is no string, but just
quarks and gluons.
This means that we should not worry of the fact that the quantum
theory of a purely
 bosonic membrane can be defined consistently only (and at most) in
$D=27$ dimensions~\cite{D27}
(as bosonic strings require $D=26$), nor of the fact  that, unlike
string theory, a membrane theory is not renormalizable by power
counting. For the same
reason, the membrane tension 
which will be introduced is not a fundamental
constant, but, as the string tension in {\em QCD}, it is in principle
derivable from the underlying theory.

In this paper we will only study black holes with zero angular
momentum. Although we believe that there is no fundamental difficulty
in generalizing our approach to Kerr black holes, in this
paper we will make  use of the technical simplifications due to
spherical symmetry.

\section{Classical  membrane dynamics}

The action of the relativistic bosonic membrane is given by 
\be\label{s1}
S=-{\cal T}\int d^3\xi \sqrt{-{\rm det}
 \left( g_{\mu\nu}\partial_i x^{\mu}\partial_j x^{\nu}\right) }
\, ,
\ee
where $\cal T$ is the membrane tension, and has dimensions of
(mass)$^3$. The membrane 
world-volume is parametrized by $\xi^i = (\tau ,\sigma_1 ,
\sigma_2)$, that is $x^{\mu}=x^{\mu}(\xi )$. The indices $i,j$ take
values $0,1,2$ and $\partial_i =\partial /\partial\xi^i$.
The metric of the
target space is $g_{\mu \nu}$, with $\mu ,\nu =0,1,2,3$ and the
signature is Minkowskian.
We  use Planck units, $\hbar =c =G=1$.
The action~(\ref{s1}) is the natural generalization of the 
relativistic action of a spinless 
 pointlike particle, which is proportional to its world-line, and
of a bosonic string, which is proportional to the world-area swept.  

This action or  modifications of it
have been considered in various physical problems.
It was first proposed by Dirac~\cite{Dir}
in 1962  as a possible model for an extended
electron.  Later it has been studied
both as a generalization of  bosonic strings and with motivations
coming from bag models of hadrons~\cite{CT,Has}. In recent years 
it has been used to investigate the bubble nucleation
process in the early Universe in inflationary 
cosmology~\cite{AS1,AS2}, while its
supersymmetric extension (``supermembrane'') received some attention
as a generalization of superstrings~\cite{BST,Duf}.

For pointlike particles and strings, the square root can be eliminated
introducing an einbein or a world-sheet metric, 
respectively; similarly, here one can introduce a world-volume
metric $h_{ij}, i,j=0,1,2$. At the classical level, the action
given in eq.~(\ref{s1}) is equivalent~\cite{HT} to the action
\be\label{s2}
S=-\frac{{\cal T}}{2}\int d^3\xi\, \sqrt{-h}
\left[ h^{ij}g_{\mu\nu} \partial_i x^{\mu}\partial_j x^{\nu}
       -1 \right]\, ,
\ee
where $h={\rm det}\, h_{ij}$.
Two differences with the  string case are noteworthy.
First, the appearance of a cosmological term in
eq.~(\ref{s2}). Second, the fact that the
metric $h_{ij}$ cannot be gauged away even at the classical level. 

The action (\ref{s2}) is invariant under general coordinate
transformations of the world-volume.
The equations of motion obtained with variation 
with respect to $h_{ij}$ are
\be\label{induced}
h_{ij}=g_{\mu\nu}\partial_i x^{\mu}\partial_j x^{\nu}\, .
\ee
Thus $h_{ij}$ is the metric induced by $g_{\mu\nu}$ on the surface of
the membrane.
Variation with respect to $x^{\mu}$ gives
\be\label{eqmoto}
\partial_i\left( \sqrt{-h} \, h^{ij}
g_{\mu\nu}\partial_j x^{\nu}\right) 
-\frac{1}{2}\sqrt{-h}\, h^{ij}g_{\rho\sigma ,\mu}
\partial_i x^{\rho}\partial_j x^{\sigma}
=0\, .
\ee
In order to find
solutions of the equations of motion with spherical symmetry one
proceeds as follows~\cite{CT}. 
One introduces ``polar fields'', i.e. a set
of three fields $r(\xi ),\theta (\xi ),\phi (\xi )$
(where $\xi$ is a shorthand for 
$(\tau,\sigma_1,\sigma_2)$) defined by
\be\label{xmu}
x^{\mu}
=(x^0,r\sin\theta \cos\phi ,
r\sin\theta \sin\phi ,
r\cos\theta )\, .
\ee
Next one fixes the gauge
\be\label{gauge}
x^0(\xi )=\tau,\hspace{5mm}	\theta (\xi )=\sigma_1,
\hspace{5mm}\phi (\xi )=\sigma_2\, .
\ee
This gauge fixing is useful also at the quantum level,
since in this gauge there is no Faddeev-Popov ghost~\cite{Flo}.
The parameters $\sigma_1,\sigma_2$ satisfy
\be
0\leq\sigma_1\leq\pi\, ,\hspace{5mm}0\leq\sigma_2\leq 2\pi\, .
\ee
One now makes the spherical 
ansatz, $r(\tau,\sigma_1,\sigma_2)=r(\tau )$. 
This reduces the problem to a single degree of freedom $r(\tau )$. Of
course, at the classical level this simply means that we are looking for
 special solutions of the equations of motion, but the solutions we get
are exact. At the quantum level, instead, restriction to membranes of
this form is a drastic approximation, analogous to the
minisuperspace approximation used in quantum cosmology.  However, in
the case of black holes, the  no-hair theorem ensures that at the
classical level the horizon has spherical symmetry, so we expect that
the approximation might be useful even in the quantum theory.

In any application, an especially important issue is the existence of
stable classical solutions.
The classical equations of motion of the 
action~(\ref{s2})  have been studied  in flat target
space, $g_{\mu\nu}=\eta_{\mu\nu}$, in refs.~\cite{CT,Sug}.
In this case, no stable classical
solution has been found. The membrane contracts under its
 self-attraction, until
it collapses.  While strings can stabilize
themselves rotating, this is not possible for a membrane in $D=4$,
since the centrifugal force will be zero on the rotation axis
(stable rotating classical solutions exist in dimensions higher than
four~\cite{KY}). Alternatively, one can study 
the equations on spaces with
non trivial topology, such as $S^2\times R^{D-2}$, where
stable solutions exist~\cite{Duf,Flo}.

In the Dirac extended electron model, instead, the self 
attraction is
balanced by the electrostatic repulsion, giving rise to a stable
solution.  However, the model is unstable under quadrupole
deformations from spherical symmetry~\cite{Has}.

The equations of motion have also been studied in the case of a
self-gravitating membrane~\cite{AS1,AS2,HOT}, in which the metric is
generated by the membrane itself, so that it is taken to be a
Reissner-Nordstrom metric outside the membrane and a flat metric
inside, with suitable junction conditions. We will instead be
interested in the motion of a membrane in a background metric which is
fixed from the beginning (it is generated by the matter which has
collapsed to form the black hole) and 
in this section we only  consider the situation
in which the back-reaction of the membrane on the metric is
neglegible. The limits of this approximation will be discussed in
sect.~3.1. 

We will find that stable  
solutions of the action given in eq.~(\ref{s1}) exist when the 
metric $g_{\mu\nu}$ is taken to be a metric with a horizon.
Let us now analyze separately the
Reissner-Nordstrom and Rindler cases.

\subsection{Reissner-Nordstrom metric}
We consider a general Reissner-Nordstrom black hole with mass
$M$ and charge $Q$.  
We use Boyer-Lindquist coordinates, which are appropriate for
a fiducial observer,  and we only
consider the metric outside the horizon. As we will see, this approach
is consistent since, for the fiducial observer, a membrane initially
outside the horizon always remains outside.

The Reissner-Nordstrom metric is given by
\be
ds^2 = -\alpha dt^2 +\alpha^{-1} dr^2 + r^2(d\theta^2+\sin^2\theta
d\phi^2)\, ,
\ee
\be
\alpha =1-\frac{2M}{r}+\frac{Q^2}{r^2}=\frac{1}{r^2}(r-r_+)(r-r_-)
\, ,
\ee
and $r_{\pm}=M\pm \sqrt{M^2-Q^2}$. The outer horizon is at $r=r_+$.
The induced metric is
computed from eq.~(\ref{induced}) and 
in the gauge $x_0(\tau ,\sigma_1,\sigma_2)=\tau$ is
\be
h_{ij}=\left( 
  \begin{array}{ccc}
     -(\alpha^2 -\dot{r}^2)/\alpha &    &         \\
                        & r^2 &                   \\   
                        &     & r^2\sin^2\theta
  \end{array}
\right)\, ,
\ee
where $\dot{r}=dr/d\tau$.
Using the ansatz~(\ref{xmu}), the  equation of
motion~(\ref{eqmoto}) with $\mu =0$ gives
 immediately
\be\label{eqrad}
\partial_0\left[ r^2\alpha 
(\alpha -\frac{\dot{r}^2}{\alpha})^{-1/2}\right] =0\, .
\ee
The equations with 
$\mu =1,2,3$ give a single second order differential equation 
\be\label{eqrad2}
r\ddot{r}+2(\alpha^2-\dot{r}^2)+\frac{r\alpha '}{2\alpha}
(\alpha^2-3\dot{r}^2)=0\, ,
\ee
where $\alpha '=d\alpha /dr$. Eq.~(\ref{eqrad}) is the first integral
of eq.~(\ref{eqrad2}),
 except that eq.~(\ref{eqrad}) also admits the
solution $\dot{r}=\ddot{r}=0$,
$r=$ arbitrary constant, while in eq.~(\ref{eqrad2}) this is a
solution only if this arbitrary constant is equal to the horizon
radius (so that $\alpha =0$). Then,
the equation governing the radial motion is 
eq.~(\ref{eqrad2}), or eq.~(\ref{eqrad})
with the solution $r=r_0$  removed unless the constant
$r_0$ is equal to $r_+$. Therefore, 
we have found that there is a stable
static solution corresponding to a spherical membrane lying on the
horizon.

Integration of eq.~(\ref{eqrad})  gives
\be\label{eqint}
\dot{r}^2 = \alpha^2 - C^2r^4\alpha^3\, ,
\ee
where $C$ is an
integration constant fixed by the initial conditions. (Note that 
timelike motions outside the horizon
satisfy the condition $\alpha^2 -\dot{r}^2 >0$).
Eq.~(\ref{eqint}) is formally identical to the motion of a 
non-relativistic point
particle with unit mass and zero energy  in the potential 
$U(r)=(-\alpha^2 + C^2r^4\alpha^3)/2 \, ,r>r_+$ (see fig.~1).
\begin{figure}[t]
\vspace{70mm}
\includegraphics{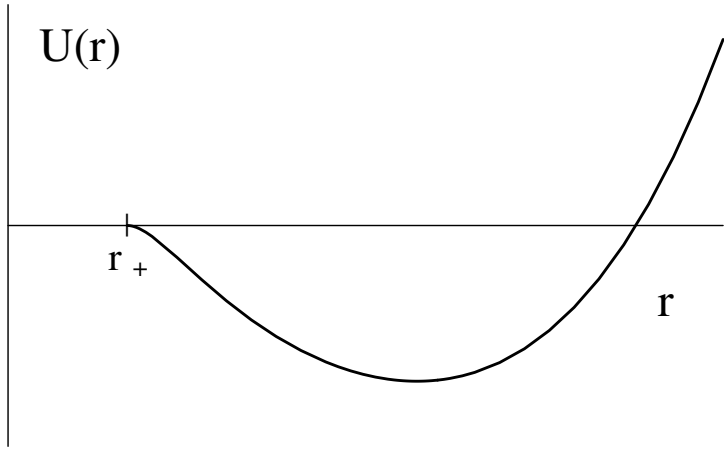}
\caption{The effective potential discussed in the text. The classical
motion of the spherical membrane is formally equivalent to the motion
of a
particle with $E=0$ in this potential.}
\end{figure}
Depending on the sign of $\dot{r}(0)$, 
the membrane can expand until it reaches a maximum value
and then recontract, or it will contract
immediately. In both cases, it
reaches asymptotically $r=r_+$, in a logarithmically divergent time
(or in a linearly divergent time for extremal black holes,
$M=Q$).
This asymptotic approach to the horizon is just what  we expect for
any object, as seen by a fiducial observer.

Note that we never need to know the metric
inside the ``nominal'' horizon, $r=r_+$. 
A membrane initially at $r>r_+$ will always remain
outside, and a fiducial observer can consistently describe the dynamics
of the membrane  without making any reference to the region $r<r_+$. 
Of course, this is not surprising; it is just what happens for any
object, pointlike or extended, from the point of view of a fiducial
observer. 

If we substitute the spherical ansatz~(\ref{xmu}) into the 
membrane action, eq.~(\ref{s2}), with the induced metric given
by eq.~(\ref{induced}) and integrate over the
 variables $\sigma_1,\sigma_2$, we find the action
\be\label{Seff}
S=-4\pi{\cal T}\int d\tau\, r^2\left( \alpha -
\frac{\dot{r}^2}{\alpha}\right)^{1/2}
\ee
The Euler-Lagrange equation of this action is  
eq.~(\ref{eqrad2}). Thus, eq.~(\ref{Seff}) is the effective action for
the radial coordinate.

It is interesting to study the stability of the solution that we have
found under non-spherical deformations. Using again the gauge fixing
(\ref{gauge}) but keeping
 $r (\tau,\sigma_1,\sigma_2)$  generic,
from eq.~(\ref{eqmoto}) with $\mu =0$ we immediately find the
more general equation
\be\label{nonsph}
\dot{r}^2(\tau ,\theta ,\phi )=
\alpha^2-C^2(\theta ,\phi )r^4\alpha^3\, ,
\ee
and $C$ is the integration constant, whose dependence on
$\theta ,\phi$ is determined by the dependence of
$r$ and $\dot{r}$
on $\theta ,\phi$ at the initial time $\tau =\tau_0$,
$r(\tau_0,\theta ,\phi )=r_0(\theta ,\phi )$.
We see that, as the horizon is approached, $\alpha\ra 0$
and   the $C$-dependent term in eq.~(\ref{nonsph}) become less and
less important. If at $\tau =0$ the membrane is sufficiently close
to the horizon, the subsequent evolution is given, for a Schwarzschild
black hole, by
\be
r(\tau,\theta ,\phi )\simeq 2M+
\left( r_0(\theta ,\phi )-2M\right) \exp \{ -\tau /(2M)\}\, ,
\ee
and any initial deviation from spherical symmetry is washed out in an
exponentially small time. This is exactly the behavior expected for a
black hole from the classical no-hair theorem.

In the above discussion, the equations of motion have been studied in
terms of a field $r(\tau ,\sigma_1,\sigma_2)$ which, by definition, is
constrained to satisfy $r\ge 0$.  Furthermore, we are only considering
the equations of motion in the region $r\ge r_+$.
When we have in mind to proceed to a
quantum field 
theory, such a field is not appropriate, since a constraint
of the form $r\ge 0$
(or $r\ge r_+$) is very difficult to implement and has quite non
trivial effects on the dynamics. We therefore introduce a
tortoise field $r_*(\xi )$, 
analogous to the standard tortoise coordinate and
defined by
\be\label{tort}
dr_*=\frac{dr}{\alpha}\, .
\ee
In particular for a
Schwarzschild black hole, with an appropriate  
choice for the integration constant
in eq.~(\ref{tort}),
\be\label{tortS}
r_*=r+2M\log\frac{r-2M}{2M}\, .
\ee
The tortoise field defined by eq.~(\ref{tort})
 ranges from $-\infty$ to $+\infty$ as $r$ goes from 
the horizon, $r=r_+$, to
infinity. We can rewrite the equations of motion in terms of the
field $r_*(\tau ,\sigma_1,\sigma_2)$, making the ansatz that
$r_*$ actually depends only on $\tau$. Eq.~(\ref{eqint}) then becomes
\be
\dot{r}_*^2=1-C^2r^4\alpha\, ,
\ee
where now $r$ is a function of $r_*$ given implicitly by
eq.~(\ref{tort}) (or by eq.~(\ref{tortS}) for a Schwarzschild 
black hole).
In terms of $r_*$ there is no static solution, since
the solution  $r=r_+$ has been mapped to  $r_*=-\infty$. 
Rather, at
large values of $\tau$ the solution is approaching $r_*=-\infty$
asymptotically. 

\subsection{Rindler metric}
We now consider the
 Rindler metric, which is the metric appropriate to
an observer with constant acceleration $g$ in flat Minkowski space.
Let  $(T,x,y,Z)$ be the Minkowski coordinates. If we define new
coordinates $t,z$ from
\begin{eqnarray}\label{zztt}
T &=&z\sinh gt\\
Z &=&z\cosh gt\nonumber\, ,
\end{eqnarray}
the flat Minkowski metric takes the form
\be
ds^2=-g^2z^2dt^2+dx^2+dy^2+dz^2\, .
\ee
This spacetime has an horizon at $z=0$. 
Besides its intrinsic interest, this metric is
the limit of the Schwarzschild metric for large black hole masses,
after an appropriate change of variables. In fact, defining 
\be
x=2M(\theta -\frac{\pi}{2})\, , \hspace{3mm}
y=2M\phi\, ,\hspace{3mm}
z=4M\sqrt{1-\frac{2M}{r}}\, ,
\ee
and identifying $g=1/(4M)$,
the Schwarzschild and Rindler metrics differ only by terms $O((z/M)^2), 
O((x/M)^2)$. The region $z\ge 0$ is mapped in the region $r\ge 2M$. The
fact that the spatial metric is flat has a number of simplifying
features which makes this metric particularly useful for our analysis.
We will consider membranes in
the region $z\ge 0$, and at the classical level we look for a solution
of the equations of motion for the four fields $x^{\mu}(\tau ,
\sigma_1,\sigma_2)$ of the form
\be\label{ansatz}
 \begin{array}{lcl}
t(\tau ,\sigma_1,\sigma_2)  =  \tau     & \hspace*{4mm} &
x(\tau ,\sigma_1,\sigma_2)  =  \sigma_1 \\
z(\tau ,\sigma_1,\sigma_2)  =  z(\tau ) & \hspace*{4mm} &
y(\tau ,\sigma_1,\sigma_2)  =  \sigma_2\, .
 \end{array}
\ee
Inserting the ansatz~(\ref{ansatz}) in the equations of motion
we find the induced metric
\be
h_{ij}=\left( 
  \begin{array}{ccc}
     -g^2z^2+\dot{z}^2  &    &                    \\
                        & 1  &                   \\   
                        &     & 1
  \end{array}
\right)\, .
\ee
The equations of motion~(\ref{eqmoto}) with $\mu =1,2$ are identically
satisfied, while the equations with $\mu =0,3$ give a 
single\footnote{As in the black hole case, the $\mu =0$ equation has
also a solution $\dot{z}=0$ which is not shared by the 
$\mu =3$ equation, unless $z=0$.} equation
for $z(\tau )$,
\be\label{eqz}
z\ddot{z}-2\dot{z}^2+g^2z^2=0\, .
\ee
As in the black hole case, we find a stable static solution on the
horizon,
$\dot{z}=\ddot{z}=z=0$. The motion with arbitrary initial conditions
can be easily studied from eq.~(\ref{eqz}), which can be integrated 
analytically. If the initial condition is
$z(0)=z_0,\dot{z}(0)=\dot{z}_0$,  and we define 
$v_0=\dot{z}_0/(gz_0)$ (which is the 
initial velocity measured in proper time
units, and satisfies $|v_0|\le 1$), 
the solution is
\be
z(\tau )=\frac{z_0}{\cosh g\tau -v_0\sinh g\tau}\, .
\ee
The qualitative features of the motion are analogous to the black hole
case. Substituting the ansatz~(\ref{ansatz}) in the membrane action
and defining ${\cal T}'={\cal T}(\int dxdy)$, we find the effective
action
\be\label{Seff2}
S=-{\cal T}'\int d\tau\, (g^2z^2-\dot{z}^2)^{1/2}\, .
\ee
The Euler-Lagrange equation of this action is
just eq.~(\ref{eqz}).

\section{Quantum membrane dynamics}
At the classical level, our approach has given satisfactory 
results. We
have found classical solutions in correspondence with horizons, and
the classical dynamics is certainly compatible with the black hole
interpretation that we are trying to develop. 
 Of course, the motivation of our investigation is to
study black holes at the quantum level.  We then proceed to 
the quantum description.

The quantization of the full membrane action~(\ref{s2}) is a very
difficult task. 
We will therefore make a drastic approximation: we
will limit ourselves to spherical membranes, $r=r(\tau )$,
in the black
hole case, and to membranes of the form $z=z(\tau )$ in Rindler space.
This approximation is analogous to the minisuperspace approximation
used in quantum cosmology. It has already been 
used in refs.~\cite{AS1,AS2} in the study of the quantum 
dynamics of self-gravitating membranes and of membranes in flat space.
Having reduced the 
number of degrees of freedom, we are giving up the
possibility of computing the black hole entropy. However, we will see
that many interesting feature are still present in this approximation. 

Our starting point,
therefore, is the action~(\ref{Seff}) in the
black hole case and the action~(\ref{Seff2}) in Rindler space.
Of course, when we proceed to a quantum 
treatement, we must also ask under
what conditions  the loop corrections to the action
(\ref{s1}) or to the action~(\ref{s2}) (which in general are not
equivalent at the quantum level) are under control. We will discuss
this point in sect.~\ref{limit}.

\subsection{Reissner-Nordstrom metric}
In this case the Lagrangian is
\be\label{Leff}
L=-{\cal T}'r^2(\alpha -\frac{\dot{r}^2}{\alpha})^{1/2}\, ,
\ee
where we have defined ${\cal T}'=4\pi {\cal T}$. To quantize the
system, we  pass to a Hamiltonian description.  One can try to
proceed defining the conjugate  momentum $p$ 
\be
p=\frac{\delta L}{\delta \dot{r}}={\cal T}'\frac{r^2\dot{r}}{\alpha}
(\alpha -\frac{\dot{r}^2}{\alpha})^{-1/2}\, .
\ee
This equation can be inverted to give $\dot{r}$ as a function of $p$,
and the Hamiltonian $H=p\dot{r}-L$ can be computed. We find 
\be\label{h1}
H=\sqrt{\alpha^2p^2+{\cal T}'^2 r^4\alpha}\, .
\ee
The square root is not a problem, since the expression inside the
square root is positive definite outside the horizon. However,  
when we go over 
to the quantum theory and impose canonical commutation
relations between $r$ and $p$, eq.~(\ref{h1}) is beset by
a very severe ordering ambiguity because of the term
$\alpha^2p^2$, and thus it is not suitable as a starting point for
quantization. Furthermore, when we have in mind that our quantum
mechanical problem is a first approximation to a field theoretical
problem, then $r$ is not the most appropriate variable since,
as we mentioned above, the field 
$r(\xi )$ is subject to the condition $r\ge 0$ by definition, 
and to the condition $r> r_+$ if we want to set up a quantum
description of the black hole which does not make reference to the
black hole interior, as 
it is implicit in the membrane approach, and such
conditions are difficult to implement in a field theory. 
Thus neither the variable
$r$ nor the Hamiltonian~(\ref{h1}) are appropriate.

Remarkably, we have 
found  that the introduction of the tortoise field,
eq.~(\ref{tort}), which solves the problem of the constraint
$r>r_+$, also gives a very natural solution of the ordering
ambiguity. 
We define the momentum conjugate to $r_*$,
\be
p_*=\frac{\delta L}{\delta \dot{r}_*}={\cal T}'
\alpha^{1/2}r^2\dot{r}_*(1-\dot{r}_*^2)^{-1/2}\, .
\ee
For the Hamiltonian, $H=p_*\dot{r}_*-L$, we get
\be\label{h2}
H=\sqrt{p_*^2+{\cal T}'^2r^4\alpha}\, ,
\ee
which does not present any ordering ambiguity.\footnote{Actually, an
ambiguity still exists connected with the arbitrary constant
in the integration of eq.~(\ref{tort}). 
 We will discuss it carefully in sect.~\ref{limit},
where we will find that it fixes the limit of validity of the
membrane description.}
In eq.~(\ref{h2})
$r$ is a function of $r_*$ given implicitly by eq.~(\ref{tort}).
We now quantize the system imposing
\be
\left[ r_*,p_*\right] =i\, 
\ee
and describe the quantum membrane with a wave function $\psi 
(r_*)$ which gives the amplitude of finding the membrane
at the given value $r_*$. The Schroedinger equation for the black hole
is then $H\psi_E =E\psi_E$ with $H$ given in eq.~(\ref{h2}). 
At the level of first quantization this is equivalent to
\be\label{scr}
\left( -\frac{d^2}{dr_*^2}+{\cal T}'^2r^4\alpha\right)\psi_E(r_*)
=E^2\psi_E(r_*)\, .
\ee
This equation can be interpreted as a Schroedinger equation in one
dimension for a particle with mass $m=1/2$ in a potential
\be\label{pot}
V(r_*)={\cal T}'^2r^4\alpha (r)\, ,
\ee
with $r=r(r_*)$ given by eq.~(\ref{tort}), and with $E^2$ playing the
role of the eigenvalue. But of course, one recognizes immediately that
the origin of eq.~(\ref{scr}) must be an equation of the Klein-Gordon
type. In fact, Aurilia and Spallucci~\cite{AS2} have derived the
analogous Klein-Gordon equation in the case of flat space. In our
derivation of the Lagrangian~(\ref{Leff})
we have chosen the gauge $x^0(\tau ,\sigma_1,\sigma_2)=\tau$. If
we do  not fix the gauge, but work with $x^0$ generical,
we find instead the Lagrangian
\be
L=-{\cal T}' r^2\left[ \alpha 
(\frac{dx^0}{d\tau})^2-\frac{\dot{r}^2}{\alpha}
\right]^{1/2}\, .
\ee
Defining the momenta conjugate to $r_*$ and $x^0$,
\begin{eqnarray}
p_0=\frac{\delta L}{\delta \dot{x}^0}&=&
-{\cal T}'r^2\alpha^{1/2}\dot{x}^0[(\dot{x}^0)^2-r_*^2]^{-1/2}\\
p_*=\frac{\delta L}{\delta \dot{r}_*}&=&
+{\cal T}'r^2\alpha^{1/2}\dot{r}_*[(\dot{x}^0)^2-r_*^2]^{-1/2}\, ,
\end{eqnarray}
we see that they satisfy the relation
$p_0^2-p_*^2={\cal T}'^2r^4\alpha$, from which it follows the
equation
\be\label{kg}
\left( -\frac{\partial^2\hspace*{1mm}}{\partial t^2}
+\frac{\partial^2\hspace*{1mm}}{\partial r_*^2}\right)
\phi (r_*,t)={\cal T}'^2r^4\alpha\phi(r_*,t)\, .
\ee
In the flat space limit, $\alpha\ra 1,r_*\ra r$, we recover 
the corresponding equation
found in ref.~\cite{AS2}.

As long as we are interested in the first quantization of the membrane
the two approaches, based respectively on the Schroedinger equation,
eq.~(\ref{scr}), or on the Klein-Gordon equation, eq.~(\ref{kg}),
 are equivalent. In the following we shall use the
approach based on the Schroedinger equation, which is intuitively
appealing.  If one studies the second quantization of the membrane
in the minisuperspace approximation
(which is beyond the scope of this paper) 
eq.~(\ref{kg}) is the correct starting point.\footnote{Depending
on the spin, the square root
can also be eliminated in favor of a Dirac equation, see
ref.~\cite{Hos} for a similar treatement in flat space. However, 
we are only considering black holes with ${\bf J}=0$.}

The potential $V(r_*)$ is plotted in fig.~2. We see that $V(r_*)$ also
illustrates nicely the classical motion. The stable classical solution
is at $r_*=-\infty$, and a membrane with given energy,
depending on whether $\dot{r}(0)$ is positive or negative, 
can expand up to
a maximum radius and recontract, or it can immediately contract and
approach  the horizon asymptotically.

\begin{figure}[t]
\vspace{70mm}
\includegraphics{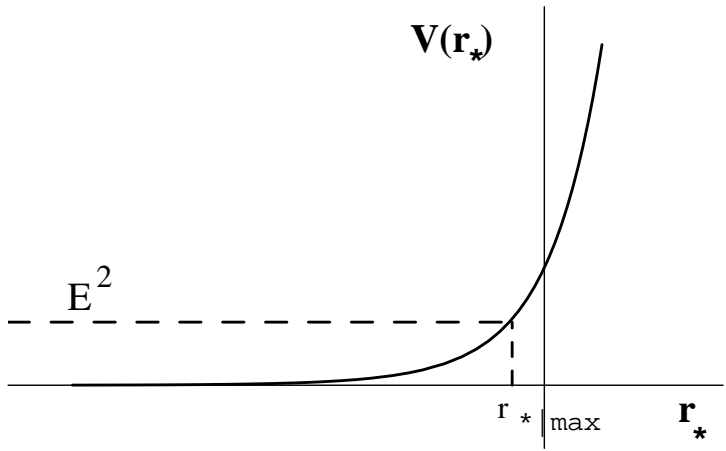}
\caption{The effective potential $V(r_*)$ vs. $r_*$. The wave function
with energy $E$ decreases exponentially for $r_*>r_{*|{\rm max}}$.}
\end{figure}

Let us now examine the qualitative features of
the Schroedinger equation in the potential $V(r_*)$. 
The fact that $V(r_*)\ra\infty$ as $r_*\ra\infty$ is very welcome. It
means that, independently of the value of the excitation energy $E$,
the wave function of the black hole vanishes exponentially for 
sufficiently large values of $r_*$. In particular, no tunneling of the
membrane to infinity is possible, which would have made our
description quite unreasonable. (Or at least very dangerous, since it
would have implied that a black hole has a
 small but  finite probability of expanding until it covers the whole
Universe!) 

The fact that $V(r_*)\ra 0$ as $r_*\ra -\infty$, and therefore the
existence of a continuous set of energy levels, is also welcome. It
means that the black hole can absorbe arbitrarily small quanta of
energy. This point, however, is susceptible of a small but important
modification. If we introduce the stretched horizon~\cite{TPM,Sus1},
we limit the range of $r_*$ to $r_*>-L$, where $L$ is a large but
finite constant. The energy levels then become discrete, although
very finely spaced. 
This point will be discussed in sect.~\ref{spectrum}. 

Since the metric $g_{\mu\nu}$ has
been fixed at the beginning, this quasi-continuum of levels 
belongs to a black hole with a given, fixed, 
mass $M$ and charge $Q$, and hence with
fixed  area of the horizon.  Thus, we consider
them as excitations of a black hole with a given area. 
Actually, an excited level with excitation energy $E$ has a larger
value of the classical horizon radius, 
 determined graphically in fig.~2 from the condition
$E^2=V(r_{*|{\rm max}})$.  Of course, if $E$ is sufficiently large
the corresponding value of $r_{*|{\rm max}}$, and hence of the horizon
area, can be made arbitrarily large. However, in this limit our
membrane approach is not self-consistent, since we have neglected any
back reaction of the membrane on the metric. Consider instead values of
the excitation energy restricted by the condition
\be\label{cutoff}
E^2\ll  {\cal T}'^2 \kappa r_+^3\, ,
\ee
where the surface gravity
of a black hole, $\kappa$, is defined as
\be
\kappa =\frac{1}{2}\frac{d\alpha}{dr}|_{r=r_+}=
\frac{r_+-r_-}{2r_+^2}\, .
\ee
Let $r_h$ be the value of $r$ corresponding to $r_{*|{\rm max}}$.
From the explicit expression of the potential, eq.~(\ref{pot}),
we get
\be
r_h\simeq r_++\frac{E^2}{2{\cal T}'^2 \kappa r_+^4}\, ,
\ee
so that $r_h-r_+\ll 1$ because of eq.~(\ref{cutoff}) and because 
$r_+\gg 1$. For the variation of the area of the horizon we get
\be
\Delta A =4\pi (r_h^2-r_+^2)\simeq
\frac{4\pi E^2}{{\cal T}'^2 \kappa r_+^3}\ll 1\, .
\ee
Thus, in this limit  the variation of the area is 
much smaller than one Planck unit and, as we will argue in
sect.~\ref{spectrum}, such a small variation is not observable.
In this domain we expect that the  approximation of
neglecting the backreaction of the membrane is
self-consistent. Furthermore, it is natural to 
 consider all membrane levels restricted by eq.~(\ref{cutoff}) as
excitations of a black hole with a given, fixed, area of the horizon.
This point will be discussed further in sect.~\ref{spectrum}.

The above results means that,  given the mass and the charge, the 
quantum state of a Reissner-Nordstrom 
black hole is not uniquely determined, contrarily
to what happens in the classical description. There is
still  a large number of microstates, corresponding to the energy
levels that we have found.
This quasi-continuum of levels is not visible in the classical
description of black holes, and the corresponding coarse-graining gives
rise to the  black hole entropy. We do not attempt to compute
the entropy from our model, since, having frozen all degrees of freedom
apart from the radial coordinate, the levels that we are discussing
are only a (very small) subset of all black hole levels. However, it
is quite clear that in a full treatement of the membrane dynamics the
number of microstates must grow exponentially with the area.

It is interesting to examine the form of the potential $V(r_*)$ for
large negative values of $r_*$.
From the definition, eq.~(\ref{tort}), one finds that
for large negative values of $r_*$ the relation between 
$r$ and $r_*$ is, for $\kappa\neq 0$,
\be
r-r_+\simeq {\rm const.}\, e^{ -2\kappa |r_*|}\, .
\ee
Then one finds immediately that
\be\label{40}
V(r_*)\simeq {\cal T}'r_+^4 (r-r_+)
\frac{d\alpha}{dr}|_{r=r_+}\simeq {\rm const.}\, 
e^{-2\kappa |r_*|}\, \hspace{10mm} (r_*\ra -\infty )\, .
\ee
Even if in eq.~(\ref{scr}) $V(r_*)$ 
formally plays the role of a potential,
dimensionally it is rather a potential squared; thus, the
``physical potential'' $V_{\rm ph}=\sqrt{V}$ 
close to the horizon is a Yukawa potential
corresponding to the exchange of a particle of mass $\kappa$.
For extremal black holes, $Q=M$, $\kappa =0$.
The relation between $r$ and $r_*$ is
\be
r_*=r+2r_+\log (\frac{r-r_+}{{\rm const.}})-\frac{r_+^2}{r-r_+}\, ,
\ee
so that as $r\ra r_+$, $r_*\simeq -r_+^2/(r-r_+)$. The potential is
\be
V(r_*)\simeq {\cal T}'^2r_+^4\frac{1}{2}(r-r_+)^2
\frac{d^2\alpha}{dr^2}|_{r=r_+}\, .
\ee
Then, for $V_{\rm ph}$ we find a Coulomb potential,
\be
V_{\rm ph}(r_*)\simeq \frac{{\rm {\cal T}'r_+^3}}{|r_*|}\, 
\hspace{10mm} (r_*\ra -\infty )\, .
\ee
The appearence of the parameter $\kappa$ 
in eq.~(\ref{40}) is quite suggestive of a
relation with Hawking radiation, since $\kappa$ is related to Hawking
temperature by $T_H=\kappa/(2\pi )$.
We will discuss this point further in
sect.~\ref{spectrum}.

\subsection{Rindler metric}
In Rindler space we start with the Lagrangian eq.~(\ref{Seff2}),
\be
L=-{\cal T}'\left( g^2z^2-\dot{z}^2\right)^{1/2}\, .
\ee
As in the black hole case, if we define a momentum $p$ conjugate to
$z$ we find an Hamiltonian, $H=pz-L$, which is beset by ordering
ambiguities: 
\be
H=gz\sqrt{p^2+{\cal T}'^2}\, .
\ee
By analogy with black holes, we define a coordinate $z_*$ from
$dz=gzdz_*$, or
\be\label{z*}
z_*=\frac{1}{g}\log\frac{z}{z_0}\, ,
\ee
where $z_0$ is an arbitrary integration constant.
Contrarily to the black hole case, the relation between
$z_*$ and $z$ can be inverted analytically, 
$z(z_*)=z_0\exp (gz_*)$. Then we define
\be
p_*=\frac{\delta L}{\delta \dot{z}_*}
\ee
and the Hamiltonian $H=p_*z_*-L$ is
\be
H=\sqrt{p_*^2+{\cal T}'^2g^2z^2}=
\sqrt{p_*^2+\mu^2\exp\{ 2gz_*\}}\,\, ,
\ee
where we have defined $\mu ={\cal T}'gz_0$. Note that the
arbitrary constant $z_0$ only appears in the combination $z_0{\cal
T}'$, and therefore
it is reabsorbed in a single constant $\mu$ which
characterizes the membrane.
This Hamiltonian does not have ordering ambiguities, and 
$z_*$ ranges  between $-\infty$ and $+\infty$.
The qualitative 
features of
the potential $V(z_*)=\mu^2\exp\{ 2gz_*\}$
are similar to the black hole case, and the same
considerations apply here. However, in this case the Schroedinger
equation can be solved exactly. The equation is
\be
\left( -\frac{d^2\hspace*{1mm}}{dz_*^2}+\mu^2
e^{ {\mbox \small 2gz_*} }\right)
\psi_E (z_*)=E^2\psi_E (z_*)\, .
\ee
This is the wave equation of Liouville quantum mechanics~\cite{DJ}.
Introducing $u=(\mu /g)\exp (gz_*)={\cal T}'z$, 
which ranges between zero and
infinity, the equation becomes
\be\label{53}
\psi_E''+\frac{1}{u}\psi_E'-(1-\frac{E^2}{g^2u^2})\psi_E=0\, ,
\ee
where the prime denotes differentiation with respect to $u$. 
Eq.~(\ref{53}) is a
Bessel equation of imaginary order. Taking into account the boundary
condition $\psi =0$ at $u=\infty$, the solution is
\be
\psi_E (u)= N(E)K_{ {\mbox\normalsize i\frac{E}{g}}}(u)
\ee
where  $K$ is the MacDonald function and
$N(E)$ is a normalization constant to be determined below.
If $u\gg E/g$, $\psi_E (u) $ decreases exponentially
\be
\psi_E (u)\simeq N(E)\sqrt{\frac{\pi}{2u}}\,e^{-u}\, ,
\ee
while close to the horizon, $u\simeq 0$, the wave function, expressed
again as a function of $z_*$, is a superposition of incoming and
reflected waves,
\be\label{56}
\psi_E (z_*)\simeq N(E)\frac{ig}{2E }
\left[ (\frac{\mu}{2g})^{iE/g}\Gamma (1-i\frac{E}{g} )
e^{ {\mbox \normalsize iEz_*} }
-(\frac{\mu}{2g})^{-iE/g}\Gamma (1+i\frac{E}{g} )
e^{ {\mbox \normalsize -iEz_*} }\right]\, .
\ee
($\Gamma$ is the Euler gamma function).
For later considerations, it is important to compute also the
normalization factor. 
If we use standard continuum energy normalization the result 
is~\cite{DJ}
\be\label{N1}
N(E )=\left(\frac{\sqrt{2}\sinh (\pi E/g )}{\pi g}\right)^{1/2}\, .
\ee
Another possible normalization is suggested by the fact that
in curved space (if $g_{00}$ is time-independent)
a continuity equation
takes the form (see e.g. ref.~\cite{TPM}, eq.~(3.46))
\be
\sqrt{-g_{00}}\,\frac{\partial j^0}{\partial\tau}=-\nabla\cdot 
(\sqrt{-g_{00}}\,{\bf j})\, ,
\ee
where $\tau$ is the proper time and $j^{\mu}=(j^0 ,{\bf j})$
is a conserved current.
The integrated form,
for $j^0 = |\psi |^2$, is
\be
\frac{\partial\hspace*{1mm}}{\partial\tau}
\int_V d^3V\,\sqrt{-g_{00}}\, |\psi |^2 =
-\oint_{\partial V} d{\bf s}\cdot {\bf j}\sqrt{-g_{00}}\, .
\ee
This suggests that another useful
normalization of the wave function, in (3+1) dimensions, is given by
\be
\int d^3x\,\sqrt{-g_{00}}\, |\psi |^2 =1\, .
\ee
(We extend the integration only to the region outside the horizon.)
In our case, the problem is one-dimensional and
$g_{00}=-g^2z^2$. Thus, we require
\be\label{norm}
\int_0^{\infty}dz\, gz|\psi |^2 =1\, .
\ee
This gives the normalization factor
\be
\tilde{N}(E )
={\cal T}'\left(\frac{2\sinh (\pi E/g )}{\pi E}\right)^{1/2}\, .
\ee
In fig.~3 we plot 
the quantity $\rho$ defined as
\be
\rho (u) =\frac{1}{{\cal T}'}\sqrt{-g_{00}}\, |\psi_E(u)|^2\,  ,
\ee
where $\psi_E$ is normalized with $\tilde{N}(E)$,
 so that
$\int_0^{\infty}du\, \rho (u)=1$.
At small values of $u$ the wave
function has an oscillatory behavior, see eq.~(\ref{56}). 
If $E\lsim g$ the amplitudes of
these oscillations are very small compared to the heighest peak, and
they are not visible in the scale of fig.~3a. For large values of $E$
they become quite visible, see fig.~3b.

\begin{figure}[ptb]
\vspace{70mm}
\includegraphics{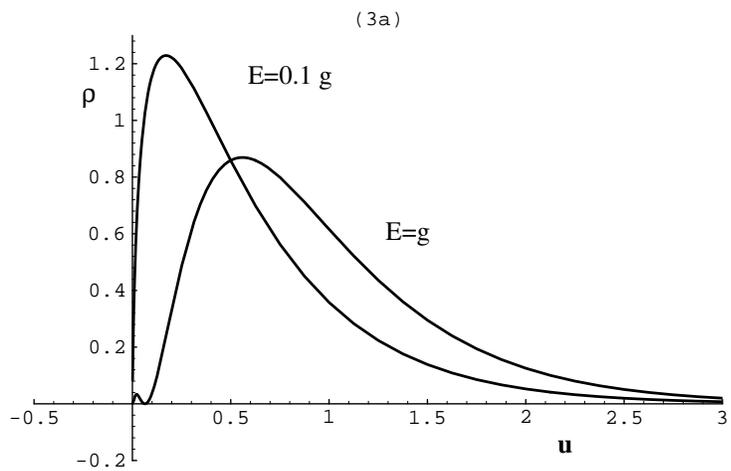}
\vspace{70mm}
\includegraphics{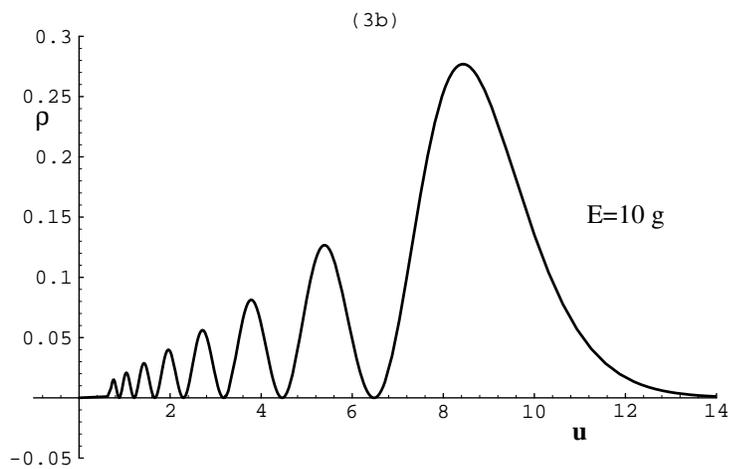}
\caption{The probability distribution of the membrane, $\rho$, vs.
$u={\cal T}'z$ for $E/g=0.1,1$ (fig.~3a) and for $E/g=10$ (fig.~3b).}
\end{figure}

\subsection{Limits of validity of the membrane 
description}\label{limit}
As we mentioned in the Introduction, the membrane approach 
should be considered as  an effective rather than a fundamental
description of black holes. We can now discuss the
domain in which we expect such 
an effective description to be adequate.

The introduction of the tortoise field has been essential in our
approach, since it provides a natural solution of the ordering
ambiguities in the Hamiltonian, and gives us a field which ranges
freely between $-\infty$ and $+\infty$. This field is defined 
by the integration  of the relation $dr=\alpha dr_*$, and a problem
arises as to the value of the integration constant. For instance, for
the Schwarzschild black hole the general solution of this 
equation can be written as
\be\label{L_0}
r_*=r+2M\log\frac{r-2M}{L_0}
\ee
where $L_0$ is an arbitrary integration
constant which is usually set equal to the
horizon radius, $2M$. Independently of $L_0$, $r_*\ra -\infty$ as
$r\ra r_+$, and $r_*\simeq r$ if $r\ra\infty$, so these boundary
conditions do not help to fix $L_0$.
All the steps leading to the Hamiltonian,
eq.~(\ref{h2}), can be repeated with $L_0$ generic, and we still obtain 
eq.~(\ref{h2}) except that now $r$ as a function of $r_*$ is given
by eq.~(\ref{L_0}) and thus depends on the arbitrary constant $L_0$.
This introduces an ambiguity in our description.
In the absence of a physical criterium for fixing this constant, we can
proceed as follows. First, since the only length scale for a
Schwarzschild black hole is given by the horizon radius, we can
set $L_0=2Ml_0$, where $l_0$ is a  numerical constant. We can now
require that the constant $l_0$ gives only subleading effects. It is
apparent from equation (\ref{L_0}) that this happens if and only if
$\log M$ is a parametrically large quantity,
so that $2M\log M$ is much larger than  $2M\log l_0$.
Thus $r(r_*)$, and therefore the Hamiltonian
$H$, do not depend on $l_0$ in a first
approximation. 
Hence, our membrane
approach gives unambiguous answers  in the domain
\be
 M\gg 1 \hspace*{10mm} {\rm (Schwarzschild)}\, .
\ee
Before proceeding to the general Reissner-Nordstrom case, it is useful
to derive again this condition in a way which is more suitable for
generalizations. We use the  variables 
$x=r/(2M), x_*=r_*/(2M), 
l_0=L_0/(2M)$; then $x=x(x_*,l_0)$ defined implicitly by
\be
x_*=x+\log\frac{x-1}{l_0}\, .
\ee
Taking the partial derivative with respect to $l_0$ we get
\be\label{x-1}
\frac{\partial x}{\partial l_0}=\frac{x-1}{xl_0}\, .
\ee
Consider the region $r\le 2M+$ const.;
if we send $M\ra\infty$ while keeping the constant
arbitrarily large but fixed, 
then in this region $x\ra 1$ and equation~(\ref{x-1}) shows that 
in this limit $x$ becomes
independent of $l_0$. Thus, if $M\gg 1$, we have 
$\partial r/\partial l_0\ll 2M$, 
which is the desired result
since (when we restore conventional units)
$2GM$ is the only length scale in the problem.

Let us repeat this analysis 
for a  non extremal Reissner-Nordstrom black hole;
the relation between $r_*$ and
$r$, with an appropriate definition of
 the integration constant $L_0$, is
\be
r_*=r+\frac{r_+^2}{r_+-r_-}\log\frac{r-r_+}{L_0}
-\frac{r_-^2}{r_+-r_-}\log\frac{r-r_-}{L_0}\, .
\ee
We define  $x=r/r_+$, $l_0=L_0/r_+$, $\rho =r_-/r_+$.
It is convenient to define also
$\varepsilon ,\delta$ from $x=1+\varepsilon,\rho=1-\delta$.
Computing $\partial x/\partial l_0$ we find
\be\label{ed}
\frac{\partial x}{\partial l_0}=
\frac{1}{l_0}
\, \frac{(2-\delta )
\varepsilon (\varepsilon +\delta )}{(1+\varepsilon )^2}\, .
\ee
In the Schwarzschild case we required
 $\partial x/\partial l_0\ll 1$ since this implies 
 $\partial r/\partial l_0\ll 2M$ and $2M$ is the only scale in the
problem. Here we also have a second scale, $\sqrt{M^2-Q^2}$. If we
ask   that $\partial r/\partial l_0$ should be
small compared to any physical
scale involved, we must require $\partial r/\partial l_0\ll
\sqrt{M^2-Q^2}$. Then we
get the condition
\be\label{condi}
\varepsilon (\varepsilon +\delta )\ll\delta\, .
\ee
For black holes far from the extremal limit, $\delta\sim 1$, we
recover the condition $\varepsilon\ll 1$, or $M\gg 1$. For
near-extremal black holes we have $\delta\simeq 2\sqrt{M^2-Q^2}/M$
and therefore $\delta$ is much smaller than
$\varepsilon\sim {\rm const.}/M$. The
condition~(\ref{condi}) then gives $\varepsilon^2\ll\delta$. In terms
of $M,Q$ this means
\be\label{xxx}
\sqrt{M^2-Q^2}\gg\frac{{\rm const.}}{M}\, .
\ee
Quite remarkably, this is the same result obtained in
ref.~\cite{PSSTW} as the limit of validity of the thermodynamical
description of black holes, in the case of zero angular
momentum.\footnote{In the more general case of non-zero angular
momentum, one finds~\cite{PSSTW} $Q^2\ra Q^2+(J^2/M^2)$ in
eq.~(\ref{xxx}). It is clear that the same result would be obtained
generalizing the membrane approach to $J\neq 0$, since 
the derivation only uses  the definition of the tortoise coordinate.}
The physical meaning  of this restriction will be examined 
from the membrane point of view after
the  discussion of  
the energy spectrum of black holes in the next section.
The region of validity of the membrane description is the area
below the curve in fig.~4, with the condition $M\gg 1$. Note that
extremal black holes are excluded by this domain of validity, although,
if $M$ is sufficiently large, we can get as close as we wish to the
extremal limit.

\begin{figure}[t]
\vspace{70mm}
\includegraphics{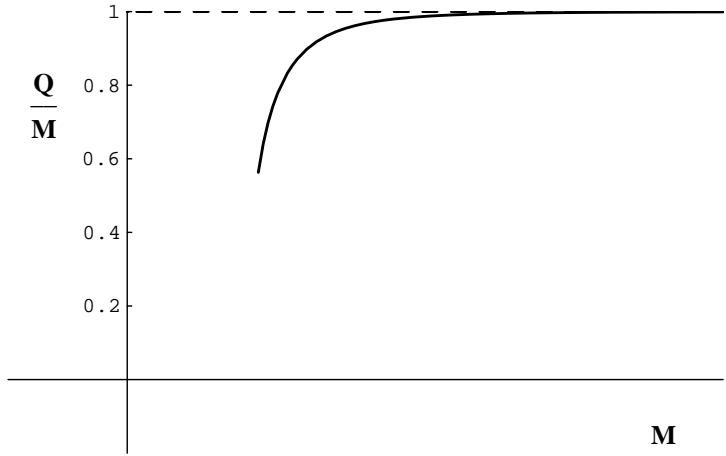}
\caption{The domain of validity of the membrane description.}
\end{figure}

A second source of worries for the consistency of our approach is due
to the fact that we are studying the quantum theory obtained applying
the minisuperspace approximation to the action, eq.~(\ref{s1}) or
eq.~(\ref{s2}), but we are neglecting loop corrections to these actions
which arise at the quantum level. However, it is easy to see what is
the parameter governing these corrections. Restoring $\hbar$,
 the parameter of the loop expansion is $\hbar /{\cal T}'$ and
therefore the expansion is under control if ${\cal T}'$ is
sufficiently large. If one computes field theoretical corrections in
the region outside the horizon, one encounters integrals over the
three dimensional space which only extends to the region $r>r_+$, so
that the relevant length scale is $r_+$. 
Since ${\cal T}'\sim ({\rm mass})^3$, 
 the loop expansion is under control if
 (setting again $\hbar =1$)
\be\label{T'r}
{\cal T}'r_+^3\gg 1\, .
\ee
This can be made more precise computing the effective static potential
for the membrane.  The computation is similar to the bosonic string 
case~\cite{LSW,AA}.
However, because of the intrinsic non-linearity
of the membrane action, it is necessary to find first a
classical solution around which to expand the partition
function. In ref.~\cite{Flo} the computation is performed considering
a membrane embedded in a space-time with topology $S^2\times R^{D-2}$,
where there are static solutions. Denoting by $R$ the radius of the
sphere $S^2$, the effective potential for the closed bosonic membrane
is, for large $R$, and in the large $D$ limit~\cite{Flo},
\be
V_{\rm eff}\simeq {\cal T}'R^2+\frac{(D-3)}{2}\,\frac{0.2651}{R}\, .
\ee
Extrapolating the result to $D=4$, we see that
the second term is small compared to the leading term if  
${\cal T}'R^3\gg 1$. This result is quite transparent physically.
The situation is completely analogous to what happens when we try to
describe the large distance effects in {\em QCD\/} in terms of a
bosonic string, so that it is useful to keep this example in mind. The
leading term is  analogous to the linearly rising potential in the
case of strings, $V_{\rm eff}\sim\sigma R$, 
which gives rise to
confinement in {\em QCD\/}. The fact that for membranes
$V_{\rm eff}\sim R^2$ while for strings it is linear in $R$ is
dictated by dimensional arguments,
since ${\cal T}'\sim ({\rm mass})^3$ while the string tension
is $\sim ({\rm mass})^2$. The $1/R$ term is generated by the
transverse fluctuations  and it is therefore proportional to the
number of transverse degrees of freedom, $D-3$ for a membrane and
$D-2$ for a string. It is also interesting to observe that, in the
large $D$ limit, the effective potential due to the action (\ref{s1})
and due to the action~(\ref{s2}) are the same 
at first order in the $1/D$ expansion~\cite{OW}. 

It is clear that in the case of membranes in a Reissner-Nordstrom
metric the role of $R$ is played by $r_+$. The leading term is the
potential $V_{\rm ph}$ that we have previously defined as the square
root of $V(r_*)$, eq.~(\ref{pot}),
and for large values of $r$ we have found
$V_{\rm ph}\sim {\cal T}'r^2$. The one-loop corrections are $\sim
1/r$ and therefore they are under control if ${\cal T}'r^3\gg 1$.
Since $r\ge r_+$, condition~(\ref{T'r}) ensures that these 
corrections are
small and that our approach is consistent.

For bosonic strings 
there is a critical distance, $\sigma R^2_c\sim (D-2)$,
below which the effective potential becomes 
 complex and the string picture breaks down~\cite{AA}. 
This is quite clear  physically, since at short distances the
quark-antiquark dynamics is governed by asymptotic freedom rather then
by confinement. Similarly  for membranes~\cite{Flo} 
there is a critical distance ${\cal T}'R_c^3\sim D-3$ below which the
approach breaks down.

We have therefore found two conditions for the consistency of the
membrane approach, eq.~(\ref{xxx}) and eq.~(\ref{T'r}). In the next
section, 
after having discussed the energy level structure of black holes, we
will be able to give a tentative estimate of the membrane tension
${\cal T}'$. Then, we will find that these two conditions are actually
one and the same.

\section{The energy spectrum of black holes}\label{spectrum}
In this section we study the level structure of black holes which is
suggested by the membrane description. Some of our considerations in
this section have only heuristic value, and the corresponding results
should only be considered   as an indication of how a black hole
might look like at the quantum level.

Let us begin by considering the
process in which a photon of energy, say, $E$=1 MeV 
falls into a
Schwarzschild black hole of mass $M$. In the classical description,
the Schwarzschild radius increases by an amount $\Delta r=2GE$ which,
for $E\sim 1$ MeV, is $\Delta r\sim 10^{-22}L_{\rm Pl}$, where $L_{\rm
Pl}$ is the Planck length. Now, it is quite obvious that such a result
does not go through at the quantum level. Most probably, no meaning
whatsoever can be attached in quantum gravity to such an
exceedingly  small
distance, nor in general to distances smaller than $L_{\rm Pl}$.
More  specifically, in the case of the black hole horizon a Gedanken
experiment~\cite{MM} indicates that the measurement of the radius of
the horizon is subject to a generalized uncertainty principle,
\be
\Delta x\gsim\frac{\hbar}{\Delta p}+{\rm const.}\, G\Delta p\, ,
\ee
which implies the existence of a minimal measurable distance on the
order of $L_{\rm Pl}$. A similar uncertainty principle was previously
found in string theory~\cite{GV,ACV1,GM,KPP}.

Because of this, it is more natural to assume that when a single
particle with energy much smaller than the Planck mass is dropped
into a black hole, the area of the horizon simply does not change. Or,
in other words, that the area is quantized, and that the  quantum
of area is of order one in Planck units. 

This argument, combined with the results of the previous section, leads
us to suggest the following pattern for the level structure of black
holes. Firstly, we 
have  a ``principal'' series of levels, corresponding to the 
area quantization. We can describe them with a ``principal quantum
number'' $n$, and at least for large $n$ we expect $A=A_0 n$, where
$A_0$ is the quantum of area, $A_0\sim L_{\rm Pl}^2$.
Each level of the principal series is 
characterized by a well-defined value of $M$ and $Q$, and to each of
these levels we can apply the membrane description developed in the
previous section, inserting for $g_{\mu\nu}$ in the action,
eq.~(\ref{s1}), the Reissner-Nordstrom metric with the given values of
$M,Q$. Thus, from  each principal level starts a continuum
(or more precisely a quasi-continuum, see below) of 
membrane levels, which are able to absorb, thermalize and reemit
quanta with energy much smaller then the Planck mass.

In our approach, we can consistently discuss only levels in which
 the excitation energy of the membrane is sufficiently small, so
that the quasi-continuum which starts at the level with principal
quantum number $n$, i.e. at $E=E_n$, ends (much) before $E=E_{n+1}$.
As discussed in sect.~3.1,
the reason for this is that we have always neglected the back-reaction
of the membrane on the metric. Of course, when the membrane absorbes an
infalling particle and jumps to an excited level, the excitation
energy $E$ gives a contribution to the metric. If the energy of the
infalling particle is sufficiently small it is legitimate to
neglect this contribution. The
approximation however breaks down when the excitation energy becomes
comparable with the spacing between principal levels.

This level structure is sketched in fig.~5. We will now discuss in
turn these two type of levels.

\begin{figure}[tbph]
\vspace{70mm}
\includegraphics{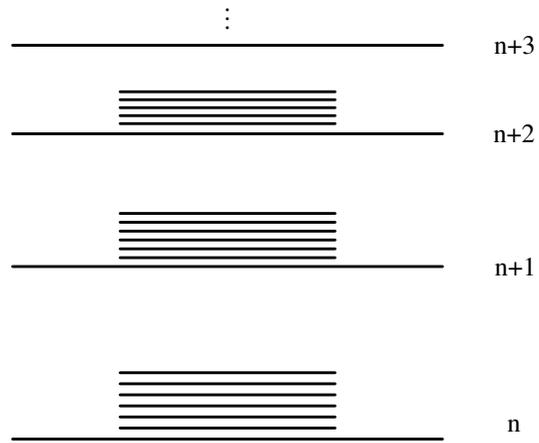}
\vspace{70mm}
\caption{The level structure of black holes.}
\end{figure}

\subsection{The principal levels}
The physical argument presented above suggests 
that the area is quantized and,
at least for large $n$,
\be
A=A_0\, n\, ,
\ee
where $A_0$ is the area quantum. It is very interesting to note, 
in this connection, that the fact that 
in quantum gravity the area of physical surfaces is
quantized 
has been suggested, from quite a different perspective, by
Ashtekar, Rovelli and Smolin~\cite{ARS} and Rovelli~\cite{Rov}
by making use of the loop representation~\cite{Loop}. The quantum of
area in this approach is
$A_0=\frac{1}{2}L_{\rm Pl}^2$, that is $A_0=1/2$ in Planck units.
In the following, however, we will keep $A_0$ generical.
For a Schwarzschild black hole, $A=16\pi M^2$.  
Setting $M=E_n$, the principal 
levels are therefore given by
\be
E_n=\left(\frac{A_0n}{16\pi}\right)^{1/2}
\ee
and the spacing
between two successive principal levels is, for large $n$,
\be
\Delta E_n =\frac{A_0}{32\pi E_n}\, .
\ee
For a  Reissner-Nordstrom black hole $A=4\pi r_+^2$ and
 one immediately finds
the more general level formula
\be\label{levels}
E_n=\sqrt{ \frac{A_0n}{16\pi}}\, 
\left( 1+\frac{4\pi Q^2}{A_0n}\right)\, .
\ee
If we set $A_0=8\pi$ we recover a level
formula which was proposed already in 1974 by Bekenstein~\cite{Bek},
who argued that classically
the irreducible mass squared of a black hole behaves as an adiabatic
invariant, and then used the Bohr-Sommerfeld quantization condition.
The spacing between successive levels is
\be\label{E1}
\Delta E_n=\frac{A_0}{8\pi}\kappa_n\, ,
\ee
\be
\kappa_n =\frac{\sqrt{E_n^2-Q^2}}{(E_n+\sqrt{E_n^2-Q^2})^2}\, .
\ee
The choice $A_0=8\pi$ has therefore an intuitive justification in the
membrane approach: we have found in sect.~3 that at the first
quantized level the  dynamics is governed by a ``physical''
potential corresponding to exchange of a Yukawa particle of mass
$\kappa_n$. This suggests that, if one develops a second quantized
formalism, the successive levels of the principal series will
correspond to levels which differ by one unit in the occupation number
of this Yukawa particle, and then $\Delta E_n=\kappa_n$.

\subsection{The membrane levels}
In sect.~3 we have found  levels corresponding to
excitations of the membrane. 
Within our model we cannot ascertain
whether these levels really form a
continuum, as we found in the above analysis,  or  they are
discrete, although very finely spaced.
The continuum of levels is  due to the
fact that the potential $V(r_*)$ goes to zero as we approach the
horizon.
From the point of view of a fiducial observer,
however, the Hawking temperature diverges on the horizon, because of
the infinite red-shift factor between the horizon and infinity. 
Thus, if $r<r_++\delta$, with $\delta$ on the order of a few Planck
lengths, we are entering a region of superplanckian temperatures, where
new physics, about which we are almost totally ignorant, comes into
play. This is in fact one of the motiviations for constructing the
stretched horizon~\cite{TPM,Sus1}.  Thus, the form 
of the potential $V(r_*)$ at a few Planck length from the nominal
horizon is not under control; or,
in other words, our model of a
black hole horizon as an infinitely thin membrane cannot be
consistently used down to distances $\sim L_{\rm Pl}$
from the nominal horizon. 

The question of whether these levels form a continuum or are discrete
is a very important one and it is strictly related to the information
loss paradox. If the levels actually form a continuum,
 the number of states in a finite volume $r_+\le r\le r_{\rm max}$
is divergent, no matter how close $r_{\rm max}$ is to $r_+$.
Correspondingly, the entropy associated to the horizon also diverges.
This divergence is the same which was found by
't~Hooft~\cite{tH} computing the entropy of a scalar field in the
vicinity of the horizon. It can be regularized  requiring that the
 scalar field vanishes within some fixed distance from
the horizon. 't~Hooft called this regulator the ``brick wall'' model.
As stressed in ref.~\cite{tH} and more recently
in ref.~\cite{Sus5}, the information loss paradox 
can be traced back to this divergence.

Since without a detailed knowledge of
physics at length scales on the order of the Planck length
we cannot decide whether the levels 
really form a continuum,
 the information loss problem actually belongs to the
domain of a full quantum theory of gravity,
as claimed in ref.~\cite{Sus3}.
However, even if strictly speaking the membrane approach, as we have
presented it, cannot give an unambiguous answer to the information loss
problem, 
it is  clear that it certainly  does not support the claim of
Hawking that unitarity is violated in black hole
evaporation~\cite{Haw}, since we
 have at our disposal a much more simple and
economical possibility, namely 
that because of the effect of the stretched
horizon the energy levels of the membrane become discrete. In the
following, we will therefore assume that this is indeed the case.

At least at a qualitative level, we can reproduce the features of the
stretched horizon imposing by hand a cutoff at $r=r_++\delta$ and
requiring $\psi (r_*)$ to be zero at the cutoff, as in ref.~\cite{tH}.
In other words, we modify the potential near the horizon, so that
it has the form depicted in fig.~6. Thus, the energy levels
become discrete, although finely spaced. 

\begin{figure}[tbph]
\vspace{70mm}
\includegraphics{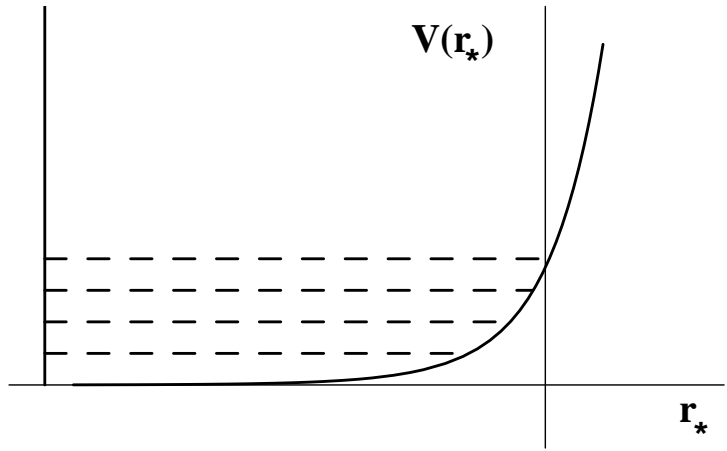}
\vspace{70mm}
\caption{The potential modified with an infinite wall at the stretched
horizon. The energy levels (dashed lines) become discrete.}
\end{figure}

However, the levels shown in fig.~6 only represent radial
excitations of the black hole, because our minisuperspace
approximation has frozen all degrees of freedom except for the radius
of the membrane.
It is quite obvious that these radial excitations
 are not relevant at
all for the description of the interaction between black holes and
infalling particles. This is also indicated by the fact that
the spacing between these levels is even larger than the spacing
between the corresponding principal levels; in fact,
approximating $V(r_*)$ with a square potential well and
remembering that the role of the eigenvalue is played by $E^2$ 
(see eq.~(\ref{scr})), we immediately find 
$E_{nk}=E_n+\left( 2\pi\kappa_n k/\log E_n\right)^{1/2}$,
where $k$ is an integer enumerating the spherical membrane levels, and
for large $n$ this gives a spacing larger than $\Delta E_n$.

On the other hand, 
when the black hole absorbes a localized particle it does not make
a transition between levels with spherical symmetry. Rather, 
classically a small bump 
is formed on the horizon
(and subsequently reabsorbed) close to the
impact point. In the quantum description, this corresponds to the
existence of a very large number of levels, 
describing states with slight deviations from spherical symmetry,
which cannot be studied within the minisuperspace approximation,
and which are much more finely spaced than the radial excitations.
Since the bump can form on any place on the surface,
the degeneration of these levels  grows with the area of the horizon.

The fact that the spectrum is not continuous anymore implies that a
black hole cannot absorb arbitrarily small amounts of energy. At first
sight this might be surprising, since if a particle moves radially 
toward the center of the black hole, we 
naively expect that it will always be absorbed.
However,  this naive expectation fails if the 
De Broglie wavelength of the particle is sufficiently large
compared to the horizon radius, since in
this case the amplitude of localizing the particle close to the horizon
is small. Thus, we expect that the separation between the membrane
levels in a state with principal quantum number $n$
will be proportional to an appropriate inverse power of $E_n$.

Another important point concerning the membrane levels is the
existence of an upper cutoff on the excitation energy of the membrane.
In sect.~3.1 we derived a bound  for the excitation energy,
 eq.~(\ref{cutoff}), from the condition that the
variation of the area of the horizon 
should be small compared to a Planck unit of
area. On the other hand, in this section we have proposed a level
formula which gives for the spacing between principal levels 
$\Delta E_n =\kappa_n$ (setting $A_0=8\pi$). It is quite natural to
assume that the maximum excitation energy of the membrane allowed by
eq.~(\ref{cutoff}) is on the order of $\Delta E_n$; in fact, when the
bound ~(\ref{cutoff}) is saturated the membrane wave function is not
yet suppressed at values of $r$ corresponding to  an area of the
horizon one Planck unit larger than the area of the nominal horizon
$4\pi r_+^2$. On the other hand, $\Delta E_n$ is just the spacing
between levels  which differ by a Planck unit of area.
This allows us to
give a tentative estimate of the value of the membrane tension.
Note that, since the membrane action, eq.~(\ref{s2}), describe the
excitations over a state with $M,Q$ given, also ${\cal T}'$, as well
as the metric $g_{\mu\nu}$, can in principle
depend on $M,Q$. Then, from  eq.~(\ref{cutoff}) and from
$\Delta E_n =\kappa_n$ we obtain
\be\label{T}
{\cal T}'^2\sim\frac{\kappa_n}{r_+^3}\, ,
\ee
where here $r_+=E_n+\sqrt{E_n^2-Q^2}$. Assuming the correctness of this
estimate, the limit of validity of the membrane description,
given in eq.~(\ref{xxx}), takes the form
\be
{\cal T}'r_+^3\gg 1\, ,
\ee
which is just the condition obtained in sect.~(\ref{limit}) requiring
that the loop corrections to the membrane action should be small. Thus,
the two consistency conditions obtained in sect.~(\ref{limit}) 
are one and the same, and are in turn
equivalent to the limits of validity of the statistical approach to
black holes, found in ref.~\cite{PSSTW};
this is quite satisfying and indicates
 that our effective
description has a good degree of inner consistency.

\section{Radiation from an accelerated 
charge and the membrane approach}
In this paper, we have mostly studied  the energy level structure of
quantum black holes. We have not discussed how  to compute
transition amplitudes between these levels. 
However, an interesting 
observation can be made in the case of Rindler space. Let us
consider, following
ref.~\cite{RW},   a massless scalar field coupled
to a source in flat Minkowski space,
\be
L=-\frac{1}{2}\partial_{\mu}\phi\partial^{\mu}\phi
+\rho\phi\, ,
\ee
where the source $\rho$ is uniformly accelerating. 
We denote by $(T,x,y,Z)$ the flat Minkowski coordinates and define
the coordinates $z,t$ as in eqs.~(\ref{zztt}).
Let us define the usual Rindler coordinate $\xi$ from $gz=\exp (g\xi )$. 
($\xi$ coincides 
with $z_*$ if we set $z_0=1/g$ in eq.~(\ref{z*})). 

The equation of motion of a uniformly accelerating particle is $\xi =$
const. In
ref.~\cite{RW} (see also~\cite{HMS}) it is considered the case
\be
\rho =\sqrt{2}q\cos (Et ) \delta (\xi )\delta (x)\delta (y)
\ee
corresponding to a uniformly accelerating charge in Minkowski space.
Furthermore, the value of the charge is oscillating with frequency $E$. 
In refs.~\cite{HMS,RW} the factor $\cos (Et)$ is inserted only as a
regulator, and the limit $E\ra 0$ is taken at the end of the
computation. However, the case $E\neq 0$ is also interesting in its
own right. The vacuum state in Minkowski space corresponds, in the
accelerating frame, to a thermal bath with temperature $T=g/(2\pi )$.
Thus, the comoving observer sees a  charge interacting
with a thermal bath, and will therefore observe spontaneous and
stimulated emission, as well as absorption. The corresponding rates
have been computed in ref.~\cite{RW}.
The total emission rate of quanta with
transverse components of the momentum $k_x,k_y$ is found to be
\be\label{dW1}
dW^{\rm em}(k_x,k_y)=\frac{q^2}{4\pi^3g}\sinh (\frac{\pi E}{g})
\left[ \frac{1}{e^{2\pi E/g}-1} +1\right]
\left| K_{{\mbox\normalsize i\frac{E}{g}}}
\left( \frac{k_{\perp}}{g}
\right)\right|^2\, ,
\ee
where $k_{\perp}^2=k_x^2+k_y^2$ and $K$ is the MacDonald function. 
The first term in the square bracket
gives the induced emission and shows that the particle sees a thermal
bath with temperature $T=g/(2\pi )$. The second term gives the
spontaneous emission rate.

The above result has a rather suggestive interpretation in the
membrane approach. From the results of section 3.2, the wave function
of the membrane corresponding to the excitation energy $E$ is
\be
\psi_E(u)=\left(\frac{\sqrt{2}\sinh (\pi E/g)}{\pi g}\right)^{1/2}
K_{{\mbox \normalsize i\frac{E}{g} }}(u)\, ,
\ee
where, in terms of $\xi$, $u=({\cal T}'/g)\exp (g\xi )$ and
we use the normalization given in eq.~(\ref{N1}). We see a
striking similarity between the spontaneous emission rate computed 
in ref.~\cite{RW} and the quantity
\begin{eqnarray}\label{dW2}
dW &=& \frac{1}{8\pi^2}\left|
\int d^3x\, \psi_E^* q\delta (x)\delta (y)
\delta (\xi )\psi_E\right|^2\, =\nonumber\\
 &=&\frac{q^2}{4\pi^3g}\sinh (\frac{\pi E}{g})
\left| K_{{\mbox\normalsize i\frac{E}{g}}}
\left( \frac{{\cal T}'}{g}
\right)\right|^2\, .
\end{eqnarray}
The integral has a natural interpretation as a matrix element of the
charge operator computed with the membrane wave function with
excitation energy $E$.
The similarity
is only approximative, since the argument of the MacDonald function in
eq.~(\ref{dW2}), after performing the integral with the help of the
Dirac delta, is ${\cal T}'/g$, while in eq.~(\ref{dW1})  it is
$k_{\perp}/g$. (Note that in the Rindler case ${\cal T}'$ has
dimensions of mass, since  ${\cal T}'={\cal T}(\int dxdy)$.) 
It is clear that such a dependence cannot be obtained
in our minisuperspace approximation, which has frozen all transverse
degrees of freedom. Still, the appearence in both equations of quite
non trivial factors as the MacDonald function of order $iE/g$ squared,
or the factor $\sinh (\pi E/g)$ is rather suggestive, and it is 
difficult to consider it as purely accidentental.

Of course, our attempt to describe quantum black holes in terms of
membranes is still in a preliminary stage and, as mentioned in the
Introduction,  we are still trying to understand what are ``the rules
of the game''. The above result might be taken as an
indication of the fact that it may be possible to describe
quantitatively the Hawking radiation in terms of transitions between
membrane levels.

\section{Conclusions}
The overall picture of black holes that emerges from our analysis
is quite in agreement with that discussed, e.g.,
in ref.~\cite{HW}. There
is a large number of low lying levels which absorb and thermalize the
energy of infalling particles. The energy is therefore emitted in a
way which, from a practical point of view, has lost any memory of  the
initial state. However, this is not different from what happens in any
ordinary macroscopic body, or in the liquid-droplet
model of the nucleus and at the fundamental level the evolution
is unitary. The membrane levels are just the ``quantum hair''  which
have been looked for (see e.g. ref.~\cite{CPW})
in order to solve the information loss problem
raised by Hawking without abandoning unitarity. 
For a given value of mass and charge, we
have found that the state of a non rotating black hole is not uniquely
determined, contrarily to what happens in classical general relativity
(no hair theorem),
but rather there exists a quasi-continuum of membrane
levels. These microstates are missed in the classical description, 
in which a state is only described with the ``macroscopic'' variables
$M,Q$ and therefore reappear in the form of entropy.

In our picture, the Hawking radiation is just the spontaneous decay
of the levels that we have found. 
The apparently thermal character of the radiation
stems from the existence of a large number of
levels, and the order of magnitude of the temperature is fixed by the
spacing between successive principal levels, i.e. $T\sim\kappa$. The
precise value of the coefficient ($1/(2\pi )$ from Hawking calculation)
cannot be computed within our minisuperspace approximation, as well
as we cannot compute the proportionality constant between the entropy
and the area. The radiation, however, only appears thermal as far as
the energy spectrum is concerned, but there are correlations between
earlier and later radiation.

Our approximations are not justified 
when the mass of the black hole becomes
of the order of the Planck mass. Thus, we cannot follow the
evaporation till the endpoint. At some point of the evaporation
process we
reach a value of $M,Q$ for which the membrane approach breaks down.
This happens at the same point when the thermodynamical description
of black holes 
also breaks down. This is understandable, since when the membrane
approach is valid we have a large number of low lying levels for each
principal level and a statistical description is therefore appropriate.
On the other side of the dividing line in fig.~4 we rather expect that 
a black hole resembles much more an elementary particle, or
a normal quantum sistem with few degrees of freedom, 
rather than a macroscopic extended object.

\vspace{15mm}

Acknowledgments. I thank Enore Guadagnini, Michele Mintchev and 
Gabriele Veneziano for useful discussions.


\begin{thebibliography}{999}
\newcommand{\pl}{{ Phys. Lett.}\ }
\newcommand{\prl}{{Phys. Rev. Lett.}\ }
\newcommand{\pr}{{ Phys. Rev.}\ }
\newcommand{\np}{{ Nucl. Phys.}\ }
\newcommand{\bb}{\bibitem}
\bb{Haw} S.W. Hawking, \pr D14 (1976) 2460.
\bb{tH} G. 't Hooft,  \np\ B256 (1985) 727.
\bb{HW} C.F.E. Holzhey and F. Wilczek, 
\np B380 (1992) 447.
\bb{Bek} J. Bekenstein, Lett. Nuovo Cimento, 11 (1974) 467.
\bb{Sus1} L. Susskind, L. Thorlacius and J.~Uglum, 
\pr D48 (1993) 3743.
\bb{TPM} K.S. Thorne, R.H. Price and D.A.~Macdonald,  
{\em ``Black Holes: The Membrane Paradigm''}, Yale Univ. Press, 1986,
and references therein.
\bb{tH2} G. 't Hooft,  \np\ B335 (1990) 138;
Physica Scripta T36 (1991) 247.
\bb{Sus2} L. Susskind, {\em ``String Theory and the Principle of Black
Hole Complementarity''}, preprint SU-ITP-93-18, July 1993,
hep-th/9307168.
\bb{Sus3} L. Susskind and L. Thorlacius, {\em ``Gedanken Experiments
Involving Black Holes''},
preprint SU-ITP-93-19, August 1993, hep-th/9308100.
\bb{Sus4} L. Susskind, {\em ``Strings, Black Holes and Lorentz
Contraction''}, preprint SU-ITP-93-21, August 1993,
hep-th/9308139.
\bb{tH3}  C.R. Stephens, G. 't~Hooft and B.F.~Whiting,
{\em ``Black hole evaporation without information loss''},
preprint THU-93/20, gr-qc/9310006.
\bb{SVV} K.~Schoutens, H.~Verlinde and E.~Verlinde,
{\em ``Black Hole Evaporation and Quantum Gravity''},
preprint CERN-TH.7142/94, hep-th/9401081.
\bb{MM} M. Maggiore, \pl B304 (1993) 65; {\em ``Black Hole
Complementarity and the Physical Origin of the Stretched Horizon''},
preprint IFUP-TH 44/93, Oct. 1993, Phys. Rev. D, to appear.
\bb{PSSTW} J.~Preskill, P.~Schwarz, A.~Shapere, S.~Trivedi and
F.~Wilczek, Mod. Phys. Lett. A6 (1991) 2353.
\bb{D27} U. Marquard and M. Scholl, \pl B227 (1989) 227;

I.~Bars, \np B343 (1990) 398.
\bb{Dir} P.A.M. Dirac, Proc. Roy. Soc. London  A 268 (1962) 57.
\bb{CT} P.A. Collins and R.W. Tucker, \np B112 (1976) 150.
\bb{Has} P. Gnadig, Z. Kunszt, P.~Hasenfratz and J.~Kuti, Ann. Phys.
116 (1978) 380.
\bb{AS1} A.~Aurilia, G.~Denardo, F.~Legovini and E.~Spallucci, 
\np B252 (1985) 523;

A. Aurilia, R.S. Kissack, R.~Mann and E.~Spallucci, 
\pr D35 (1987) 2961.
\bb{AS2} A.~Aurilia and E.~Spallucci, \pl B251 (1990) 39.
\bb{BST} J. Hughes, J. Liu and J.~Polchinski, \pl
B180 (1986) 370;

E. Bergshoeff, E. Sezgin and P.K. Townsend, \pl B189 (1987)
75.

\bb{Duf} M.J. Duff, T. Inami, C.N.~Pope, E.~Sezgin and K.S.~Stelle,
\np B297 (1988) 515.
\bb{HT} P.S. Howe and R.W. Tucker, J. Phys. A 10 (1977) L155.
\bb{Flo} E.G. Floratos, \pl B220 (1989) 61;

E.G. Floratos and G.K. Leontaris, \pl B220 (1989) 65.
\bb{Sug} A. Sugamoto, \np B215 (1983) 381.
\bb{KY} K. Kikkawa and M. Yamasaki, Prog. Theor. Phys. 76 (1986) 1379.
\bb{HOT} D.H. Hartley, M. \"{O}nder and R.W. Tucker,
Class. Quantum Grav. 6 (1989) 1301.
\bb{Hos} C. Ho and Y. Hosotani, \prl
60 (1988) 885.
\bb{DJ} E.~D'Hoker and R.~Jackiw, \pr D26 (1982) 3517, appendix~B.
\bb{LSW} M. L\"{u}scher, K.~Symanzik and P.~Weisz, \np
B173 (1980) 365.

\bb{AA} O. Alvarez, \pr D24 (1981) 440;

J.F. Arvis, \pl B127 (1983) 106.
\bb{OW} S.D. Odintsov and D.L. Wiltshire, Class. Quantum Grav.
7 (1990) 1499.
\bb{GV} G. Veneziano, Europhys. Lett. 2 (1986) 199; Proc. of Texas 
Superstring Workshop (1989).

D. Gross, Proc. of ICHEP, Munich (1988).
\bb{ACV1} D. Amati, M. Ciafaloni and G.~Veneziano, 
\pl B197 (1987) 81, B216 (1989) 41;
Int.~J.~Mod. Phys. A3 (1988) 1615; \np B347 (1990) 530.
\bb{GM} D.J. Gross and P.F. Mende, \pl B197 (1987) 129; \np B303 (1988) 407.
\bb{KPP} K. Konishi, G. Paffuti and P.~Provero, \pl B234 (1990) 276;

R. Guida, K. Konishi and P.~Provero, Mod. Phys. Lett. A6 (1991) 1487.

\bb{ARS} A. Ashtekar, C. Rovelli and L.~Smolin, 
\prl 69 (1992) 237.
\bb{Rov} C. Rovelli, \np B405 (1993) 797.
\bb{Loop} C. Rovelli and L. Smolin, \prl 61 (1988) 1155;

A. Ashtekar, {\em ``Non-perturbative canonical gravity''},
World Scientific, Singapore 1991, and references therein.
\bb{Sus5} L.~Susskind, {\em ``Some speculations about black hole
entropy in string theory''}, preprint RU-93-44,
hep-th/9309145.

\bb{RW} H. Ren and E.J. Weinberg,  
``Radiation from a moving Scalar Source'', 
preprint CU-TP-592, hep-th/9312038.
\bb{HMS} A. Higuchi, G.E.A. Matsas and D.~Sudarsky,
\pr D46 (1992) 3450.
\bb{CPW} S. Coleman, J. Preskill and F. Wilczek, \np B378 (1992) 175.
\end{thebibliography}
\end{document}